\def\etal{{\it et al.\/}}

\def\ie{{\it i.e.\/}}

\def\Pac{{Paczy\'{n}ski\/}}

\documentstyle[aasms]{article}
\tighten

\begin{document}

\title{On the acceleration of Ultra High Energy Cosmic Rays in Gamma
Ray Bursts}
\author{Mario Vietri}
\affil{Osservatorio Astronomico di Roma \\ 00040 Monte Porzio Catone (Roma),
Italy \\ E--mail: vietri@coma.mporzio.astro.it}

\begin{abstract}
UHECRs are roughly isotropic and attain very large energies, $E \gtrsim
3\times 10^{20} \; eV$. Conventional models fail to explain both
facts. I show here that acceleration of UHECRs in GRBs satisfies both
observational constraints. Using M\'esz\'aros and Rees' (1994) model of
GRBs as due to hyperrelativistic shocks, I show that the highest energies
that can be attained thusly are $E\simeq 10^{20}\;\theta^{-5/3} n_1^{-5/6}\;
eV$, explaining the energy of the
Bird \etal\/ (1995) event even without beaming.
The traditional photopion catastrophe affecting UHECR acceleration in AGNs
is circumvented. An order of magnitude estimate
shows that the total energy flux of UHECRs at the Earth is also
correctly reproduced. A test of the model based upon the UHECRs' distribution
upon the plane of the sky is briefly discussed.
\end{abstract}
\keywords{acceleration of particles -- gamma-rays: bursts}

\section{Introduction}
Ultra High Energy Cosmic Rays (UHECRs, $E > 3\times 10^{18} eV$, the `ankle')
show a flatter spectrum than Cosmic Rays (CRs) just below the ankle. They have
energies extending up to an observed  maximum of $E_{max} = 3\times 10^{20}\;
eV$ (Bird \etal, 1995), are thought to be predominantly protons (Bird \etal,
1994) and arrive from directions roughly isotropic on the plane of the sky.
Since expected angular deflections for highly energetic protons are less than
$10^\circ$ (Sigl, Schramm, Bhattacharjee 1994, SSB from now on),
this rough isotropy is thought to reflect a (rough) intrinsic
isotropy of their sites of production.

The conventional acceleration mechanism invoked so far is first order
Fermi (1949) acceleration at shocks (Axford, Lee and Skadron 1977, Blandford
and Ostriker 1978, Bell 1978, Krymsky 1977). Acceleration sites must be located
within the finite range ($\lesssim 100 \; Mpc$, SSB) over which UHECRs
can travel before their energy is significantly degraded by energy losses
due to photopion and photoelectron production off CMBR photons.
However, all acceleration sites proposed  so far
(see SSB for a beautiful review) fail on {\it two} counts.
First, they cannot attain the highest energies observed so far, falling
short by at least one order of magnitude. Second, they cannot reproduce
the rough isotropy of the directions of arrival, since they explain
all UHECRs as coming from the handful of peculiar objects (AGNs,
radiogalaxies, the Virgo Cluster and so on) that can be found within
the finite range mentioned above.
Even more frustrating is that no prospective candidate for an acceleration
site can be located within a suitable error box around the direction
of arrival of the $E= 3\times 10^ {20} \; eV$ event (SSB, Elbert and Sommers
1995).

We should thus be scouting around for a class of objects that is roughly
isotropically distributed, and where sufficient amounts of concentrated
energy are available to accelerate UHECRs. One such category, so far
unexplored, is gamma--ray bursts (GRBs, see \Pac , 1993 for a review).
I shall consider only cosmological models of GRBs,
with a total energy release of $E_ {GRB} \simeq  10^ {51} \; erg$.
I shall need in the following discussion no detailed property of the
mechanism proposed to explain the energy injection mechanism, but I shall
need the details of the hydrodymanical expansion of the fireball
leading to the GRB. In particular, the really attractive feature
that I shall try to exploit in the following is the suggestion by
M\'esz\'aros and Rees (1994, MR from now on) that hyperrelativistic
shocks (whether due to the impact of different parts of the same flow,
endowed with different Lorenz bulk factors, or to the impact of the flow on
the surrounding interstellar medium) are responsible for GRBs.

In Section 2, I shall give two generic arguments ({\it i.e.},
independent of the actual acceleration mechanism) in favor of
GRBs as sites for the acceleration of UHECRs. Those features of the
hydrodynamics of GRBs which are relevant to the problem at hand, are
briefly reviewed in Section 3, which is entirely based upon the results
of M\'esz\'aros and Rees (1994, and references therein). In section 4, I
describe the acceleration process in this
model for the evolution of the fireball.
In particular, in Section 4.1 I discuss qualitatively
two acceleration mechanisms, and I compute in subsections 4.2 and 4.3
the highest energies that UHECRs can attain. A mixed bag of limitations and
caveats are discussed in Sections 5. I discuss the results in Section
6, and summarize them in Section 7.
{}

\section{Gamma--ray bursts as accelerators of UHECRs}

There are two arguments that make GRBs appealing, the first one being
a numerological coincidence. The total energy of
UHECRs striking the Earth can be estimated as $5\times 10^{-13} \; erg\;
s^{-1} \; cm^{-2} \; sr^{-1}$. This has been obtained by taking the
UHECRs' spectrum as $N(E) \propto E^{-2.7}$, with the normalization
coming from Hillas (1984). This local flux must be compared with that
released in the form of UHECRs by GRBs. Since UHECRs have a finite range
($\leq 100 \; Mpc$, SSB), I consider only the nearby GRBs, which
occur at the rate of $\dot{n}_P = 30 \; yr^{-1} Gpc^{-3}$ (\Pac\/ 1993),
each releasing about $E_{UHECR} = 10^{51} \; erg$ in the form of UHECRs.
The total energy striking the Earth (if the sources are uniformly
distributed, and are all standard candles) is thus $\dot{n}_P D_m E_{UHECR}/
8\pi = 4\times 10^{-13} \; erg \; s^{-1}
\; cm^{-2} \; sr^{-1}$, independent of beaming. Here I took $D_m = 100 \; Mpc$
as the maximum range for $E = 3\times 10^{18} \; eV$, the threshold
energy above which I computed the observed total energy flux of UHECRs
striking the Earth. I have assumed
a kind of equipartition, such that the energy released by the GRB in the form
of photons equals that in the form of UHECRs, probably a not too bad
assumption, given the large relative velocities freely available in
GRBs. It also agrees with the well--known high efficiency of particle
acceleration at strong shocks (V\"olk, Drury and McKenzie 1984).
The above coincidence of theoretical and observed fluxes is quite striking.
It is, to the best of my knowledge, the first time that the amplitude of the
UHECRs' flux at Earth has been `explained'.

The second argument that makes fireballs attractive is that they can act as
hyperrelativistic ping--pong bats with respect to CRs.
GRBs are known to be highly super--Eddington: in fact, they show substructure
in their pulse profile on a scale $\simeq 1 \; ms$, implying source sizes
$< 3\times 10^7 \; cm$, which are, at most, the Schwarzschild radii of
$100 \; M_\odot$ objects, for which the Eddington luminosity is $\simeq
10^ {40} \; erg \; s^ {-1} $. Since GRB luminosities are
$10^ {51} \; erg \; s^ {-1} $ if at cosmological distances for typical
burst durations of $1 \; s$, clearly they are super--Eddington. It seems
likely thus that radiation pressure drives a relativistic expansion, with
large Lorenz factors $\gamma_b$. This conclusion is strengthened by the
following argument. GRB spectra extend at least to $100 \; MeV$.
Even more astounding is the recent discovery (Hurley \etal, 1994)
that at least one of these spectra extends to $18 \; GeV$.
If this energy derived from thermal energy, then electron/positron pair
creation
would easily make the burst optically thick, and the emerging spectrum
would be thermal (\Pac , 1986). Since instead GRB spectra are known to be
much broader than black--body, $\gamma_b$ must refer to
bulk motion kinetic energy (\Pac, 1993). The hydrodynamic evolution of
a fireball shows that a shell containing all of the fireball's
energy and baryon content is accelerated to relativistic Lorenz factors
$\gamma_b$, and remains thin through most of its evolution.
A magnetic field $B$, initially at
equipartition, decreases as the shell expands; when the fireball ejecta
impact upon the surrounding interstellar medium, the shocked shell
magnetic field is revived to equipartition values. Thus,
the expanding shell contains a significant, random field $B$ which might
deflect incoming cosmic rays backward. We would then have a very efficient,
first order Fermi acceleration,
for consider a highly relativistic particle of Lorenz factor $\gamma_{CR} $
moving radially inward toward the origin, which, after having entered
the shell,  may be deflected backwards by a magnetic
irregularity which is comoving with the expanding flow. In the reference frame
of the flow, it can easily be shown that relativistic
composition of velocities implies that the cosmic ray has Lorenz factor $\simeq
2 \gamma_b \gamma_ {CR} $ both before and after being turned backwards, but
after the deflection the cosmic ray, now moving radially outwards, has, with
respect to the lab frame, a Lorenz factor $\simeq 4 \gamma_b^2 \gamma_ {CR}$:
\ie , the single backward deflection has increased its energy by a factor
$4\gamma_b^2 \simeq 10^5$, for the values $\gamma_b \simeq 10^2 - 10^3$
favored by MR. Thus, if a few such cycles can be
achieved, it seems possible that CRs
can be accelerated up to energies of $3 \times 10^ {20} \; eV$,
the highest energy event detected by the Fly's Eye.

These generic remarks, which are meant to be independent of the specific
acceleration mechanism, make GRBs palatable. Below I shall discuss  the
physical processes involved in the deflection/acceleration mechanism.

\section{The hydrodynamics of GRBs}

I review here briefly some properties of GRBs which are relevant to
the acceleration of UHECRs. This subsection is entirely based upon the
results of M\'esz\'aros, Laguna and Rees (1993, MLR from now on).
Rees and M\'esz\'aros (1992) have proposed that thermalization of the spectrum
of the GRB can be avoided by introducing a slight contamination of
baryons within the original fireball, parametrized by the parameter $\eta$
\begin{equation}
\label{eta}
\eta \equiv \frac{E_0}{M c^2}
\end{equation}
where $M$ is the total baryon mass. MLR showed that, for a certain range
of $\eta$, only a fraction of the initial energy is still in the form
of radiation at the moment in which the expanding fireball becomes optically
thin, the rest having been converted into kinetic energy of the baryons.
This kinetic energy becomes available for radiation in the optically thin
regime if the baryons' directed kinetic energy can be suitably randomized,
either by collisions with the external interstellar medium, or with slower
or faster portions of the relativistic flow itself (M\'esz\'aros and Rees
1993). Since it seems rather difficult and contrived to produce a fireball
with very little baryon contamination, and since this is the favoured
model of MR, below I shall concentrate exclusively
on high--load fireballs, \ie, those with
\begin{equation}
\label{etarange}
1< \eta < \Gamma_m \equiv 3.3\times 10^5 E_{51}^{1/3} r_6^{-2/3}\;,
\end{equation}
where $E_{51}$ is the total energy release $E_0$ in the fireball in units of
$10^{51} \; erg$, and $r_6$ is the radius $r_0$ in which such energy is
released
in units of $10^6\; cm$. For these values of $\eta$, MLR showed that the
baryons can be accelerated up to $\gamma_b = \eta$, ending up with just about
all the initial energy $E_0$ in kinetic form, while only a fraction
$\eta/\Gamma_m$ is released in photons in the first, miniburst. After an
initial period (which is of no interest to us) in which the fireball
accelerates from rest up to $\gamma_b = \eta$, a period of free expansion
follows. In the lab frame, the energy in this phase is concentrated in a
thin slab, initially
of roughly constant thickness $\delta\!r \simeq r_0$, equal to
the radius of the region in which the initial energy deposition took place.
The free expansion begins when the shell has reached in the lab frame
a radius $r \simeq r_s$ such that \begin{equation}
\label{rs}
r_s =  \eta \theta^{-1} r_0
\end{equation}
where $\theta$ is the beaming semi-opening angle, $\theta=1$ corresponding
to isotropic emission,
and ends when, in the lab frame, the shell starts expanding, linearly with
radius $r$.  This occurs for $r\gtrsim r_b$, with
\begin{equation}
\label{rb}
r_b = \eta^2 r_0 \;.
\end{equation}
I shall need in the next section the expression for several quantities in
the frame comoving with the shell, expressed in terms of the lab frame shell
distance from the origin, $r$. MLR give, for the comoving radiative
energy and temperature
\begin{equation}
\label{ET}
\left(\frac{E}{E_0}\right) = \left(\frac{T}{T_0}\right) = \left\{
\begin{array}{ll}
\eta^{-1/3} \theta^{-2/3} (r_0/r)^{2/3} & r_s < r < r_b \\
\eta^{1/3} \theta^{-2/3} (r_0/r) &  r > r_b
\end{array} \right.
\end{equation}
Here $T_0 = 4\times 10^{11}\; K\; E_{51}^{1/4} r_6^{-3/4}$ is the initial
temperature. They also give the shell's thickness in the comoving frame
$\delta\!r$ as
\begin{equation}
\label{deltar}
\delta\!r = \left \{
\begin{array}{ll}
\eta r_0 & r_s < r < r_b \\
r/\eta & r > r_b
\end{array} \right.
\end{equation}
MLR also argue that a large class of phenomena leads to the growth of a
magnetic field $B$, in near equipartition with the initial radiative energy.
They deduce that the comoving value of $B$ decreases according to
\begin{equation}
\label{bfield}
\frac{B}{B_0} = \left \{
\begin{array}{ll}
\eta^{-2/3} \theta^{-4/3} (r_0/r)^{4/3} & r_s < r < r_b \\
\eta^{2/3} \theta^{-4/3} (r_0/r)^{2} &  r > r_b \\
\end{array} \right.
\end{equation}
and the initial, equipartition field $B_0$ is given by
\begin{equation}
\label{b0}
B_0 = 10^{17} \; E_{51}^{1/2} r_6^{-3/2} \xi^{1/2} \; G
\end{equation}
where $\xi$ is the ratio of magnetic to radiative energy, and measures
the departure of $B$ from equipartition (which occurs for $\xi = 1$).

The release of the energy stored as baryons' kinetic energy, in the form
of the $\gamma$--ray flash occurs either by collisions of the parts of
the fireball's endowed with different values of the Lorenz factor
(M\'esz\'aros and Rees 1993), or by impact of the fireball ejecta
on the interstellar medium. The two different mechanisms have been
incorporated into a single, coherent picture (MR)
to explain the exceptional event observed by Hurley \etal, 1994.
I shall consider only the second, latter shock, and shall speak loosely of
shocked shells to refer to this scenario only. The physics of these
collisions (M\'esz\'aros and Rees 1993,
MLR) is similar to that of the development of SN shocks. A collisionless shock
moves into the unshocked material with Lorenz factor $\sqrt{2} \gamma_{b}$,
while a reverse compression wave moves into the shell to be decelerated,
eventually steepening into a mild shock with a modest Lorenz factor, $\simeq
2$. The material immediately behind the forward shock is heated to a
thermal Lorenz factor $\approx \gamma_{b}$. In the shocked shell, several
phenomena (MLR) can revive the magnetic field to equipartition.

The impact of the fireball ejecta occurs at a deceleration
radius
\begin{equation}
\label{rd}
r_d = 10^{18} \; \theta^{-2/3} E_{51}^{1/3} n_1^{-1/3} \eta^{-2/3} \; cm
\end{equation}
where the particle density per $cm^3$ in the ISM is $n_1$. The shocked
shell has thickness
\begin{equation}
\label{shellthickness}
r_{sh} = 10^{18} \; \theta^{-2/3} E_{51}^{1/3} n_1^{-1/3} \eta^{-5/3} \; cm
\end{equation}
and I scale the intershell magnetic field $B$ with the value reached by
equipartition with the baryons' energy density, which is the same, whether
the material has been shocked by the forward or reverse shock. I find
\begin{equation}
\label{bintershell}
B = 0.5 \; n_1^{-1/2} \eta \theta^{-1} \xi^{1/2} \; G\; .
\end{equation}
where $\xi$ again parametrizes departures from equipartition ($\xi = 1$). MLR
also consider the possibility that $B$ does not, after all reach equipartition.
In that case, it is limited from below by the fossil magnetic field
(Eq. \ref{bfield}) formed at the outset of the expansion. For simplicity,
they show that this corresponds to setting $\xi \approx 4\times 10^{-3}$ in Eq.
\ref{bintershell}.

\section{The acceleration of UHECRs}

\subsection{The acceleration mechanism}

There are actually two distinct acceleration mechanisms that I want to
consider. The first, possibly less interesting one is as follows. The
freely expanding shell can deflect backward some CRs which, after the
scattering, as discussed in Section 2, shall be boosted up in energy
by a factor $4 \gamma^2_b \simeq 10^5$. In this mechanism, the shell
simply scoops up whatever CRs are floating around, acting rather passively.
Clearly, the flux estimate of Section 2 does not apply to this mechanism.
However, it is nonetheless interesting because, after the explosion that
shall eventually lead to the GRB,
since the shell expansion is hyperrelativistic,
business outside goes on as usual, and it is known by direct observations that
Cygnus X--3 (Samorski and Stamm 1983), the Crab pulsar
(Dzikowski \etal , 1981, 1983), Hercules X-1 and Vela X-1 (see Protheroe
1994 for references and a critical, but optimistic review)  produce
cosmic rays with $E \approx 10^ {16} \; eV$. The presence of high--energy
cosmic rays around normal neutron stars also agrees with theoretical
prejudice, since most conventional estimates (Hillas 1984, SSB) of the
highest energy attained by cosmic rays around neutron stars agree on the
value $E \simeq 10^{17} \; eV$. Thus, if any of these CRs were to
move backward and land on the expanding shell, it would be boosted up
to very high energies. This mechanism can be extended to a self--consistent
one, by considering the possibility that suprathermal particles
(possibly preexisting the fireball explosion) are accelerated
by repeatedly scattering off two subportions of the fireball, each
having given rise to a relativistic shell. It is easy to see that the
highest energy thusly attainable is the same as in the scoop--up model.
The details of the scattering mechanism limit
the highest energy attainable: this is discussed in Section 4.2.

The second, more appealing acceleration mechanism occurs because the
MR model for GRBs explicitly predicts the existence
of hyperrelativistic shocks, because of the impact of the expanding flow on the
surrounding interstellar medium. Since such events are seen to
have a complex structure, it seems likely that in any case we are
witnessing the collision of several subportions of the flow with each other,
so that the GRB consists of several sections of converging flows, alternatively
separated by a reverse shock, a contact discontinuity, and a forward shock.
The contact discontinuity is likely to be unstable, leading to mixing
of material on either side of it, so that particles from the extreme
Boltzmann tail of Lorenz factor $\gtrsim \gamma_b$ are injected over the
whole shocked shell.
The suprathermal particles shuffle between
the pre--shock shell and the post--shock shell, each time being scattered
by the shell's magnetic field (Eq. \ref{bfield}). At every such loop
their energy is boosted up by a factor $4 \gamma_{b}^2 \approx 10^5$ (see
the argument of Section 2 and, most importantly, the simulations of Quenby and
Lieu 1989). Even if we take the energy of suprathermal protons to be as low
as $\gamma_{b}$ ({\it i.e.}, equal to that of thermal particles),
we see that after two cycles the protons have energy $16 \gamma_{b}^5
\gtrsim 10^{11}$, roughly enough to account for the highest energy event
observed so far. Thus, the proposed mechanism is essentially a two--cycle
Fermi--Bell acceleration. This mechanism is of course limited by the ability
of the shell to deflect UHECRs backward.
Limitations arise because the shocked shell has both finite
thickness and finite lifetime, both to be considered in the following Section
4.3, together with other sundry problems. It is however clear at this point
already that the highest energy that can be attained is reached when the
shocked shell is largest, and the whole shocked shell thickness
is equal to one mean free path for the UHECR's deflection. At the same time,
another subportion of the flow has collided with this shell, and it
too is completely shocked. Then the CR shuffles back and forth between two
shells of thickness given by Eq. \ref{shellthickness}
each with a revived magnetic field (Eq. \ref{bintershell}). The computation
below shall use this argument. Also, I should add that this is a
coherent mechanism, one {\it i.e.\/} where the injection and acceleration
of suprathermal particles are all due to the same physical environment, so
that the estimates of the efficiency of the generation of cosmic rays
(V\"olk, Drury and McKenzie 1984) and the equipartition argument of
Section 2 apply.

I now compute the maximum energy to which CRs can be
accelerated. Such upper limit exists because the shells have a finite
thickness and magnetic field $B$ so that an UHECR's mean free path
to diffusion, being a multiple of the particle's gyroradius, cannot
exceed the shell's thickness; outside the shell the magnetic field
is in fact negligible with respect to that inside the shell. The computation
shall be carried out in the shell's reference frame, and then the
limiting energy shall be transformed to the lab frame. A few important comments
on the physics of the scattering of CRs are deferred to the Section 5,
in order not to obstruct the flow of the argument.

\subsection{The interaction of CRs with the expanding fireball}

I begin with determining the optical depth to photopion destruction of
an incoming CR, as a function of the shell's expansion radius. Using
$\sigma_\pi = 10^{-28} \; cm^2$ as the relevant cross--section (Caldwell
\etal, 1978), I find (with the computation done in the comoving frame)
\begin{equation}
\tau_\pi = n_\gamma \sigma_\pi \delta\!r = \frac{E}{k T \; 4\pi r^2 \delta\!r}
\sigma_\pi \delta\!r
\end{equation}
and, using Eq. \ref{ET}, I find
\begin{equation}
\label{taupi}
\tau_\pi = 1.3\times 10^{14} \left(\frac{r_0}{r}\right)^2 E_{51}^{3/4}
r_6^{-5/4} \; .
\end{equation}
Optical thinness to photopion destruction is then achieved beyond $r_\pi$,
\begin{equation}
\label{rpi}
r_\pi = 1.1\times 10^7 r_0 E_{51}^{3/8} r_6^{-5/8} \;.
\end{equation}
This occurs for $r< r_b$ or for $r> r_b$, depending upon whether $\eta>
\eta_l$, or $\eta < \eta_l$ respectively, where
\begin{equation}
\label{etal}
\eta_l = 3.3\times 10^3 E_{51}^{3/16} r_6^{-5/16}\;.
\end{equation}

Once the shell has become optically thin to photopion destruction, every
incoming relativistic CR finds itself in a magnetic field, with respect to
which it is super--Alfv\'enic. In fact, in the comoving frame where baryons
are locally at rest, the Alfv\'en speed $V_A$, given by $V_A^2 = B^2/4\pi
\rho$,
is determined by the baryons' rest--mass density $\rho_b$. This is because,
as shown by MLR, for $r > r_s$, the internal energy
in the shell comoving frame is subrelativistic (this is really the meaning of
the parameter $r_s$), and the same applies for an equipartition $B$--field.
Thus any CR with $\gamma_{CR} \gtrsim 2$ is clearly super--Alfv\'enic.
Explicitly, we have
\begin{equation}
\label{va}
\frac{V_A^2}{c^2} = \frac{B^2}{4\pi\rho_b c^2} = \frac{E}{M c^2} =
\eta \frac{\xi E}{E_0} = \xi \left\{ \begin{array}{ll}
\eta^{2/3}\theta^{-2/3} \left(\frac{r_0}{r}\right)^{2/3} & r< r_b \\
\eta^{4/3} \theta^{-2/3} \frac{r_0}{r} & r > r_b \end{array} \right.
\end{equation}
where I used Eq. \ref{ET}.
The CR being super--Alfv\'enic, the excitation of the usual helical modes that
lead to scattering is possible before $r \simeq r_\pi$: in fact, $V_A/c < 1$
for $r > r_s$. In particular at the time of thinning to photopion destruction,
$r/r_0$ is given by Eq. \ref{rpi}, so that I obtain
\begin{equation}
\frac{V_A}{c} = 0.09 E_{51}^{-1/16} r_6^{5/48} \xi^{1/2}
\left\{ \begin{array}{ll}
\left(\frac{\eta}{\eta_l}\right)^{1/3} & \Gamma_m > \eta > \eta_l \\
\left(\frac{\eta}{\eta_l}\right)^{2/3} & \eta < \eta_l \end{array} \right.
\end{equation}
For low--load fireballs, $\eta > \Gamma_m$, I derive the maximum Alfv\'en
speed at photopion thinning,
\begin{equation}
\frac{V_A}{c} = 0.09 E_{51}^{-1/16} r_6^{5/48} \xi^{1/2} \left(
\frac{\Gamma_m}{\eta_l} \right)^{1/3}
= 0.4 E_{51}^{-17/144} r_6^{-1/72} \xi^{1/2}
\end{equation}
independent of $\eta$, which shows the shell to be subalfv\'enic for any load.

The deflection mean free path of relativistic cosmic rays is generally taken,
by analogy with the interplanetary medium, to be a multiple $g$ of the
particle's gyroradius $r_L$, where $g \simeq 40$ for relativistic shocks
(Quenby and Lieu 1989). Thus, in the comoving frame, the deflection of a
CR with energy $E = 10^{15} E_{15} \; eV$ requires that the shell's thickness
(Eq. \ref{deltar}) is at least $g$ times its gyroradius, \ie, $g r_L \leq
\delta\!r$. Using
\begin{equation}
\label{rl}
r_L = 1\;  pc \frac{E_{15}}{B/(1\mu G)} \;,
\end{equation}
and Eq. \ref{bfield} and \ref{deltar} we get an upper limit to the
comoving energy of the CR which can be deflected backwards. I find
\begin{equation}
\label{emax1}
\frac{E_{max}}{10^{15}\xi^{1/2}\; eV}  =
1.0\times 10^9 \eta^{-7/3}
E_{51}^{1/2} r_6^{-1/2} \theta^{-4/3}
\left\{
\begin{array}{ll} \left(\frac{r_b}{r}\right)^{4/3} & r < r_b \\
\frac{r_b}{r} & r > r_b \\
\end{array} \right.
\end{equation}
This gives the highest CR energy that can be deflected backwards by the shell
after the thinning radius $r_\pi$. Before that time, a smaller range of CRs
could still be deflected, provided the total optical depth to photopion
destruction seen along their walk through the shell, $\simeq g r_L$, is less
than unity; in other words, provided $\tau_\pi g r_L /\delta\!r \leq 1$. This,
and Eq. \ref{taupi}, \ref{deltar}, \ref{rl} give
\begin{equation}
\label{emax2}
\frac{E_{max}}{10^{15}\xi^{1/2}\; eV}  =
6.4\times 10^{-6} \eta^{5/3}
E_{51}^{-1/4} r_6^{3/4} \theta^{-4/3}
\left\{
\begin{array}{ll} \left(\frac{r}{r_b}\right)^{2/3} & r < r_b \\
\frac{r}{r_b} & r > r_b \\
\end{array} \right.
\end{equation}

{}From Eq. \ref{emax1} and \ref{emax2} it can be seen that the highest
energy in the comoving shell is attained just at the moment of optical
thinning $r_\pi$, and, in the lab frame, this maximum energy is given by
\begin{equation}
\label{EMAX}
E_{max} \simeq 2 \eta E_{max}(r_\pi) =
4\times 10^{19} \; eV\;
E_{51}^{1/4} r_6^{-1/12} \theta^{-4/3} \xi^{1/2}
\left\{
\begin{array}{ll} \left(\frac{\eta}{\eta_l}\right)^{4/3} & \eta > \eta_l \\
\left(\frac{\eta}{\eta_l}\right)^{2/3} & \eta < \eta_l \\
\end{array} \right.
\end{equation}
This peak on the maximum energy for $r = r_\pi$ is due to the competition
between photopion destruction, which dominates at small radii, and the
decrease in the comoving magnetic field, which makes the CR's gyroradius
increase beyond the shell's thickness.

Another loss mechanism that is potentially important is due to synchrotron
radiation, which damps the UHECR energy on a timescale $t_s = 1\; yr\;
(10^{20}\; eV/E)(1\; G/B)^2$. This is to be compared with the time that
the UHECR spends within the shell, where the field is high, $t_d = 2g r_L/c$.
Using the formulas above, I find
\begin{equation}
\frac{t_s}{t_d} = 141 \;
\theta^4 \xi^{-3/2} \left(\frac{r}{r_\pi} \right)^4
\end{equation}
for the highest energy CR that can be reflected at shell radius $r$.
At $r=r_\pi$, we have $t_s/t_d \gg 1$, and even more so later on,
which shows synchrotron losses to be negligible, when no beaming is present.
Alternatively, the synchrotron cooling time provides a lower limit
to the amount of beaming consistent with the acceleration of UHECRs,
since we must have $t_s/t_d \geq 1$. Supposing $\theta < 1$, there is a
characteristic radius $r_\theta$ such that synchrotron losses do not damp
the CR, given by
\begin{equation}
\label{rtheta}
\frac{r_\theta}{r_\pi} = \frac{0.3 \xi^{3/8}}{\theta} \; .
\end{equation}
If $r_\theta < r_\pi$, the highest energy attainable is given by Eq.
\ref{EMAX}, otherwise, going back to Eq. \ref{emax1}, I find
\begin{equation}
\label{EMAXtheta}
E_{max}^{(\theta)} = 2\eta E_{max}(r_\theta) = 9\times 10^{19} \; eV \;
\xi^{1/2} E_{51}^{1/2} r_6^{-1/2} \left\{ \begin{array}{ll}
(\frac{\eta}{\eta_\theta})^{4/3} & \eta > \frac{\eta_\theta}{\theta^{1/2}} \\
(\frac{\eta}{\eta_\theta})^{2/3}\theta^{-1/3}
& \eta < \frac{\eta_\theta}{\theta^{1/2}}
\end{array} \right.
\end{equation}
where I have conveniently defined $\eta_\theta \equiv (0.3 \xi^{3/8})^{1/2}
\eta_l$. The above equation is the result we were searching for. It
shows that UHECRs are easily produced in the free expansion phase of
the events leading to a GRB, for a moderate beaming. In fact, using
MR's favoured value $\eta \simeq 10^3$, I find
that a beaming of $\theta \simeq 0.01 \simeq 1^\circ$ is necessary to
produce the highest energies observed to date, $\simeq 3\times 10^{20}\; eV$.

It is interestng to notice that, since $\gamma_b \leq \Gamma_m \simeq
4\times 10^5$ in GRBs, whether high or low load, (Shemi and Piran 1990, MLR),
there is an absolute, universal maximum to the CR energy in this acceleration
mechanism, given by
\begin{equation}
\label{eassoluto}
E_{abs} = 1.9\times 10^{22} E_{51}^{4/9} r_6^{-5/9} \theta^{-4/3} \xi^{1/2}
\; eV \;.
\end{equation}

\subsection{The acceleration of UHECRs in the shocked shell}

The acceleration of UHECRs in the shocked shell after the impact of the
fireball ejecta on the interstellar medium is subject to the same physics
as in the previous paragraph. The Alfv\'en speed is, as remarked at the end
of Section 3.1, below the speed of light and
the photopion catastrophe is no longer a
problem. In fact, the energy radiated is now the whole kinetic energy of
the baryons, so that the optical depth to photopion destruction is
\begin{equation}
\tau_\pi = \frac{E \sigma_\pi}{\epsilon \; 4\pi r_d^2} = 1 \times
\left(\frac{100\; eV}{\epsilon}\right) E_{51}^{1/3} \theta^{4/3}
\left(\frac{\eta}{1000}\right)^{4/3}
\end{equation}
where $\epsilon$ is the average photon energy in the lab frame. If all
of the energy were released by synchrotron losses, $\epsilon$ would be
in the X--ray (MLR), giving $\tau_\pi \simeq 0.1$. However, most of the
losses occur by Inverse Compton (MLR) in the $\gamma$--ray band, giving
$\tau_\pi \ll 1$.

Proceeding as in the previous subsection, and equating $g$ times the
Larmor radius to the shell thickness I find the maximum energy in the
comoving frame
\begin{equation}
\label{emax1shockedshell}
\frac{E_{max}^{(com.sh)}}{10^{15} eV \xi^{1/2}} =
5000 \; \theta^{-5/3} \eta^{-2/3} E_{51}^{1/3} n_1^{-5/6}
\end{equation}
and, in the lab frame,
\begin{equation}
\label{EMAXshockedshell}
E_{max}^{(sh)} = 2\eta E_{max}^{(com.sh)} = 10^{19} \;
\theta^{-5/3} \eta^{1/3} E_{51}^{1/3} n_1^{-5/6} \xi^{1/2} \; eV\; .
\end{equation}

Synchrotron cooling is not a limiting factor, leading to a very weak
upper limit on $n_1$. I find here
\begin{equation}
\label{timeratio}
\frac{t_s}{t_d} = 900 \; n_1^{4/3} \theta^{13/3} E_{51}^{1/3} \xi^{-3/2}\;.
\end{equation}
In order to account for the Bird \etal, 1995, event, we must have
\begin{equation}
\theta^{-5/3} \eta^{1/3} E_{51}^{1/3} n_1^{-5/6} \xi^{1/2} > 30
\end{equation}
which gives
\begin{equation}
\frac{t_s}{t_d} < 0.1 \; n_1^{-5/6} \eta^{13/15} E_{51}^{13/15} \xi^{-1/5}
\end{equation}
and, since it is necessary that $t_s > t_d$,
\begin{equation}
n_1 < 0.08 \eta^{26/25} E_{51}^{26/25} \xi^{-1/5} \; cm^{-3} \; .
\end{equation}
For $\eta \simeq 10^3$, the above limit becomes $n_1 \lesssim 100 \; cm^{-3}$,
which is a very weak limit.

For the MR's favoured value $\eta \simeq 10^3$ I
find
\begin{equation}
\label{EMAXshockedshellbeaming}
E_{max}^{(sh)} = 10^{20} E_{51}^{1/3} n_1^{-5/6} \theta^{-5/3} \xi^{1/2} \; eV
\end{equation}
which has a sufficiently steep dependence on $\theta$ and $n_1$ to
accommodate the energy of the $3\times 10^{20} \; eV$ event without
any trouble, even in the case of no beaming. In particular, it seems
obvious that energies up to $\simeq 10^{22} \; eV$ are quite compatible
with reasonable values of $n_1$. This is so especially because
at least one (Narayan, \Pac\/ and Piran 1992) of the mechanisms proposed to
explain the initial release of energy predicts such events to occur outside
galaxies, and all (Usov 1992 and 1994, Thompson and Duncan 1993) use pulsars,
which are {\it not} confined to the Galactic disk: $n_1$ is accordingly
reduced.
Then values as low as $\xi \approx 10^{-3}$ are compatible with the energy of
the Bird \etal\/ (1995) event. Such low value of $\xi$ also allows the
possibility that the shell's magnetic field necessary to accelerate UHECRs
is provided completely by the fossil magnetic field (Eq. \ref{bfield}): it
shall be remembered from the end of Section 3 that this field
corresponds to $\xi \approx 4\times 10^{-3}$.

Eq. \ref{EMAXshockedshellbeaming} is the major result of the paper.

$E_{max}^{(sh)} $ is subject to an absolute upper limit since $\eta \leq
\Gamma_m$, just like the limit for the freely--expanding fireball,
derived in the previous section. I find
\begin{equation}
E_{abs}^{(sh)} = 7\times 10^{20} E_{51}^{4/9} r_6^{-2/9}
n_1^{-5/6} \xi^{1/2} \theta^{-5/3} \; eV \;.
\end{equation}

Lastly, it should be mentioned that the shocked shell has a finite lifetime,
but that this provides no limitation on the maximum energy. In fact, the
acceleration process has, as a bottleneck, the time that the UHECR spends
on its last trip before being scattered for the last time, because its
mean free path increases with energy. Thus, the acceleration occurs
(in the shell frame) on a timescale $\approx g r_L /c = \delta\!r /c$ which
equals the light shell--crossing time. This is of course the shortest
timescale on which the GRB can be generated, and thus the shock lasts
at least as long as this (MLR).

\section{Caveats}

The main emphasis of this paper is on computing the highest energies that can
be attained by UHECRs in GRBs, and in fending off the most obvious loss
mechanisms, synchrotron and photopion. This explains the cavalier treatment
reserved to relativistic shock acceleration of UHECRs. Below I try to
make this treatment plausible.

Acceleration of CRs at relativistic shocks has been studied by several authors,
both in the test particle regime (Peacock 1981, Kirk and Schneider 1987,
Quenby and Lieu 1989) and in the nonlinear regime (Bell 1987, Jones and
Ellison 1991), and even in oblique shocks (Kirk and Heavens 1989,
Ballard and Heavens 1991). In the above, I basically used a test particle
approximation, which could fail if the modification of the shock structure
due to the inclusion of CRs were such as to decrease the relative speed of the
incoming and outgoing streams. However, simulations by Bell (1987) and
Jones and Ellison (1991) clearly display `thin' shocks even in the nonlinear
regime, and Bell (1987) shows that the detailed shock structure is irrelevant
for the highest rigidities.

In the previous section, I have assumed that the cosmic ray velocity is
subject only to pitch--angle scattering, and that it is reversed in each
scattering, which may seem unrealistic. Quenby and Lieu (1989) argue,
on the basis of the analogy with the interplanetary medium,
that scattering is essentially isotropic, and trajectory integrations
for the fully relativistic case
(Moussas, Quenby, and Valdes--Galicia 1987, Valdes--Galicia, Moussas and
Quenby 1992) have shown that scattering
occurs through large pitch--angle changes, with $\delta(\cos\theta) \simeq
0.5-1$. From this they deduce, through
their numerical simulations, that the energy is increased by a factor
$\gamma_b^2$ per cycle, when the proper average over all cosmic
rays' velocity directions is taken. In relativistic shocks the diffusion
approximation breaks down because the cosmic rays' velocities are not
isotropically distributed, but are instead strongly peaked toward the
radial direction (see Fig. 1 of Quenby and Lieu 1989). Since I too assumed
a radially peaked  velocity distribution, the
energy increase $\gamma_b^2$ applies to my case as well, and the difference
with my previous computation, $4\gamma_b^2$, is all due to the
substitution of backward--forward scattering with isotropic
(although large pitch--angle) scattering. This argument thus validates
my use of forward--backward scattering, the only ensuing error being a
modest factor $4$ in the energy increase per scattering.

The special relativistic Alfv\'en speed is given by
\begin{equation}
\label{relativisticalfvenspeed}
\frac{V_A^2}{c^2} \equiv \frac{B^2/4\pi}{4\epsilon_b /3 + B^2/4\pi}
\end{equation}
where $\epsilon_b$ is the (relativistic) baryons' energy density, and the
factor
$4/3$ becomes 1 in the non--relativistic limit. Scaling to equipartition
values I find $V_A/c \approx \xi^{1/2}/2$. I argued above (see discussion after
Eq. \ref{EMAXshockedshellbeaming}) that even $\xi \approx 10^{-3}$ is
acceptable, yielding $V_A/c \approx 0.01$. However, for equipartition
values the Alfv\'en speed is $\approx c/2$, more than in all simulations
mentioned above. Unquestionably, in this limit I am stretching the
applicability of the usual turbulence
arguments to the boundary of the non--relativistic regime. The most critical
limitation arises in the assumed, phenomenological link between the
mean free path to CR scattering, $\lambda$, and its gyroradius, $\lambda
\approx g r_L = 40 r_L$ (Quenby and Lieu 1989). Still, one should keep in
mind that, in the model of the previous section, matter shocked by the
reverse shock, in the shell's reference frame, is barely relativistic,
roughly as assumed by Quenby and Lieu (1989), and that relativistic effects
for $V_A/c \approx 0.5$ do not appear so extreme to force one
to abandon the previous estimate (Eq. \ref{EMAXshockedshellbeaming}).

Next, one may wonder whether
sufficient strong turbulence is present to ensure scattering of UHECRs in
the shocked shell. I have two arguments about this. First,
it seems quite likely that, given the large velocities and energies available,
very strong magnetic turbulence can be generated behind the shocks. Second,
this magnetic turbulence may have been observed already, albeit indirectly.
It is well--known, in fact, that GRBs are extremely rich in substructure
down to a scale of $\approx 1 \; ms$. In MR's model,
since the GRB arises from synchro--Compton radiation, this substructure
may be interpreted as inhomogeneities of the magnetic field: where the
magnetic field is strongest, the generation of synchrotron radiation is
more effective, and its conversion into $\gamma$--ray radiation leads to
the local peaks in the observed time structure of the GRB. Whether this
substructure corresponds to small--scale shocks, instabilities, or nonlinear
wave effects, it must correspond to inhomogeneities in the magnetic
field on the corresponding wavelengths. As a matter of fact, this argument
is similar to that usually made (Quenby and Lieu 1989) to justify the
presence of large--scale inhomogeneities in the magnetic field strength in
the jets of radiogalaxies. There, however, synchrotron radiation emission
is directly observable, while in GRBs it must be inferred from the
$\gamma$--ray observations.

The magnetic field associated with the shell is of course impossible
to predict theoretically. The only guide we can invoke is the
interstellar medium, where the magnetic field is well--known to
have achieved equipartition. On the other hand, the initial mass motions
must be so violent, that any combination of compression, shearing, turbulent
dynamo, magnetic buoyancy (Usov 1992 and 1994, Narayan, \Pac, Piran 1992,
Thompson and
Duncan 1993) are likely to lead to near equipartition values. The same comments
apply of course to the shell after it has started to decelerate, so that
the achievement of approximate equipartition seems to me by far the most
reasonable assumption. It should be noted, however, that MLR also consider the
more extreme case of `magnetic dominance', in which the magnetic energy
is in equipartition not just with the energy to be released eventually
in the GRB, but with the binding energy of the object leading to the
fireball. In this case the frozen--in magnetic field would a factor of
$30$ higher than in Eq. \ref{b0}, in which case the estimeates of the
maximum energy (Eq. \ref{EMAXtheta}) would be an underestimate by the
same factor of $30$.

\section{Discussion}

I discuss briefly here some consequences of the hypothesis that UHECRs are
connected with GRBs.

Can we be sure that most UHECRs are protons, rather than nuclei with
higher charges, as suggested by observational evidence (Bird \etal, 1994)?
If GRBs are extragalactic, the gyroradii
of iron nuclei in the intergalactic field ($\simeq 10^{-9} G$, SSB),
are of order $1 Mpc$, for a $10^{20}\; eV$ particle. Thus, they are confined
to large bubbles surrounding their sites of production, and rather far from us.
Then, diffusion in the Galactic magnetic field, in which their gyroradius
is $1 \; kpc$ for the same energy, tends to bar them access to the inner
regions of the Galaxy where we are observing them, and carry them outwards.
Thus, if GRBs accelerate UHECRs, the problem of their composition is
automatically solved, even neglecting photodestruction of heavy nuclei
at the acceleration sites.

One should not expect a close temporal association between UHECRs and GRBs.
Naively, one might think that, since the UHECRs' speed differs from $c$
by one part in $\gamma_{CR}^2 \gtrsim 10^{19}$, UHECRs should trail GRBs by
less than $10^{-3} \; s$, even if they arrive from about $100 \; Mpc$ away.
This would lead to the expectation that, every time we see an UHECR,
$\gamma$--ray satellites should observe a GRB. However, UHECRs are bent
along their path by the intergalactic magnetic field by approximately
$10^\circ$ (SSB), leading to a path longer by $\simeq 10^{15} \; s$, thus
washing away any correlation with GRBs.

The expected angular distribution of UHECRs is isotropic by construction
(because such are the GRBs, Meegan \etal, 1992),
and an absolutely unavoidable consequence of the model. Hopefully, it ought to
become testable with the construction of the Giant Airshower Detectors.
Minor departures from exact isotropy are however expected.
If GRBs are cosmologically distributed, the expected dipole
is of order $1-2\times 10^{-2}$, but most importantly it is oriented
in the direction of the Sun's peculiar motion with respect to the frame of the
CMBR $(l,b) = 264^\circ .7, 48^\circ .2$ (Maoz 1994).
However, in the cosmological case the expected dipole for UHECRs would depart
from that of GRBs because energy losses prevent the arrival of UHECRs
from distances $\gtrsim 100 \; Mpc$ (SSB). This is especially interesting
because it is exactly in this region (Scaramella, Vettolani and Zamorani 1994)
that the Local Group's peculiar velocity forms. In particular, this seems
interesting especially in connection with the well--known anisotropy of
UHECRs (Hillas 1984), which has often been ascribed to the contribution of
the Virgo Cluster, which is also known to contribute greatly to the formation
of the Local Group's peculiar velocity. Furthermore, UHECRs have a
direct cosmological application in that, by restricting attention to higher
and higher energy bins, we can select to look at closer and closer distances,
and thus ought to be able to see the dipole moment fade away. These
anisotropies are currently being studied and the results shall be published
elsewhere.

There is very little that can be said about the UHECRs' spectrum, at this
stage. In the test particle case in relativistic shocks (Kirk and Schneider
1987), the CRs' spectrum is harder than in non--relativistic shocks.
Non--linear effects can modify this conclusion, even though it is not clear
in which direction: Bell (1987) argues that, if injection is limited to
$\approx 10^{-4}$ of the background number density, the test particle spectral
shape is valid, while the non--linear computations of Jones and Ellison (1991)
find a softer spectrum than for the test particle case. How each of
these computation would be modified by a nearly relativistic Alfv\'en speed
is not known. This spectrum should also be convolved with the steepening
induced by photopion and photoelectron destruction induced by
propagation over distances $\gtrsim 10 \; Mpc$.

\section{Summary}

This paper is essentially about a coincidence. Conventional sites for the
the production of UHECRs fail to explain both the highest energies observed
so far, and the rough isotropy of their directions of arrival. It is
suggestive that there is one class of objects, those that produce
GRBs, which can remedy {\it both} these defects.

The directions of arrival of UHECRs in this model are isotropic
because such are GRBs (Meegan \etal, 1992). Two acceleration mechanisms
are qualitatively described in Section 4.1.
The highest energies that CRs can attain by bouncing off fireballs are
given by $ E \simeq 9\times 10^{19} \theta^{-1/3}\; eV$ when the fireball is in
the phase of free expansion (see Eq. \ref{EMAXtheta}). When, however,
we consider as an acceleration mechanism the
conventional Fermi acceleration at shocks, and the shocks are the
highly relativistic ones invoked by MR to explain
GRBs, then the highest energies that CRs can attain, are given by
$E\simeq 10^{20} \; \theta^{-5/3} n_1^{-5/6} \xi^{1/2} \; eV$ (Eq.
\ref{EMAXshockedshellbeaming}). In the first case (no shocks),
some beaming is necessary to explain the highest energies observed to
date (Bird \etal, 1995), while in the second one a proper choice of
$n_1$ is sufficient, thus doing without beaming. If the UHECRs are
generated at the same time as GRBs, and
if some kind of equipartition between the various forms of
energy losses is achieved in nearby ($\lesssim 100 \; Mpc$) GRBs,
then the total flux of UHECRs at the Earth is correctly predicted by an
order of magnitude estimate (see Section 2). Lastly,
the photopion catastrophe has been shown to be irrelevant after the
fireball has expanded beyond a radius $r_\pi$ (Eq. \ref{rpi}). I have argued
that a test based upon the distribution of UHECRs' directions of
arrival is feasible, and that no conclusion about the spectral shape is
possible at the moment.

Thanks for helpful discussions are due to R. Scaramella and E. Pesce, and
especially to B. Paczy\'nski.

\end{document}